\documentclass[twocolumn]{aastex62}

\graphicspath{{./}{figures/}}

\begin{document}

\title{The first detection of a low-frequency turnover in nonthermal emission from the jet of a young star}

\correspondingauthor{Anton Feeney-Johansson}
\email{antonfj@cp.dias.ie}


\author{Anton Feeney-Johansson}
\affiliation{Dublin Institute for Advanced Studies, Astronomy \& Astrophysics Section, 31 Fitzwilliam Place, Dublin 2, D02 XF86, Ireland}
\affiliation{School of Physics, Trinity College, Dublin 2, Ireland}

\author{Simon J. D. Purser}
\affiliation{Dublin Institute for Advanced Studies, Astronomy \& Astrophysics Section, 31 Fitzwilliam Place, Dublin 2, D02 XF86, Ireland}

\author{Tom P. Ray}
\affiliation{Dublin Institute for Advanced Studies, Astronomy \& Astrophysics Section, 31 Fitzwilliam Place, Dublin 2, D02 XF86, Ireland}
\affiliation{School of Physics, Trinity College, Dublin 2, Ireland}

\author{Jochen Eisl\"offel}
\affiliation{Th\"uringer Landessternwarte Tautenburg, Sternwarte 5, D-07778, Tautenburg, Germany}

\author{Matthias Hoeft}
\affiliation{Th\"uringer Landessternwarte Tautenburg, Sternwarte 5, D-07778, Tautenburg, Germany}

\author{Alexander Drabent}
\affiliation{Th\"uringer Landessternwarte Tautenburg, Sternwarte 5, D-07778, Tautenburg, Germany}

\author{Rachael E. Ainsworth}
\affiliation{Jodrell Bank Centre for Astrophysics, Alan Turing Building, School of Physics and Astronomy, \\
The University of Manchester, Oxford Road, Manchester M13 9PL, UK}



\begin{abstract}

Radio emission in jets from young stellar objects (YSOs) in the form of nonthermal emission has been seen toward several YSOs. Thought to be synchrotron emission from strong shocks in the jet, it could provide valuable information about the magnetic field in the jet. Here we report on the detection of synchrotron emission in two emission knots in the jet of the low-mass YSO DG Tau A at 152 MHz using the Low-Frequency Array (LOFAR), the first time nonthermal emission has been observed in a YSO jet at such low frequencies. In one of the knots, a low-frequency turnover in its spectrum is clearly seen compared to higher frequencies. This is the first time such a turnover has been seen in nonthermal emission in a YSO jet. We consider several possible mechanisms for the turnover and fit models for each of these to the spectrum. Based on the physical parameters predicted by each model, the Razin effect appears to be the most likely explanation for the turnover. From the Razin effect fit, we can obtain an estimate for the magnetic field strength within the emission knot of $\sim 20\ \mu \mathrm{G}$. If the Razin effect is the correct mechanism, this is the first time the magnetic field strength along a YSO jet has been measured based on a low-frequency turnover in nonthermal emission.
\end{abstract}

\keywords{Classical T Tauri stars (252) --- Non-thermal radiation sources (1119) ---  Star formation (1569) --- Stellar jets (1607) --- Stellar magnetic fields (1610)}


\section{Introduction} \label{sec:intro}
Young stars are associated with powerful jets which carry away mass and angular momentum from the system \citep{Frank2014}. Though the exact collimation and launching mechanism for the jets is uncertain, it is generally agreed that it involves a magnetic field originating in the star or the accretion disk \citep{Pudritz1983a, Shu1994}. Material from the accretion disk is lifted and centrifugally accelerated along the magnetic field lines. The jets are then also collimated by the magnetic field due to ``hoop stresses'' from the toroidal component of the field. However, despite their importance to the origin of the jet, so far magnetic fields in jets are poorly understood \citep{Ray2009} and measurements of the field strength are scarce.

Radio emission from YSOs has generally been observed to have spectral indices in the range $ -0.1 < \alpha < 1$ \citep{Anglada1996}, where the flux density $S_{\nu}$ at frequency $\nu$, $S_{\nu} \propto \nu ^{\alpha} $. This is characteristic of thermal bremsstrahlung emission and indicative of partially ionized outflows \citep{Reynolds1986}, where the bulk of the emission comes from the base of the ionized jet.

In contrast, nonthermal emission from YSO jets, characterised by negative spectral indices, is not very well studied even though it could provide valuable information about the poorly studied magnetic field \citep{Ray2009}. This is mainly because of the high sensitivities required to observe the normally weak nonthermal emission. Recently however, nonthermal emission has been observed towards several YSOs, generally in radio knots in the jet, which show negative spectral indices \citep{Anglada2018}. Its origin is thought to be synchrotron emission from relativistic electrons, accelerated either in shocks caused by the collision of the material within the jet with the dense gas in the surrounding molecular cloud, or in internal shocks within the jet due to variable ejection rates \citep{Padovani2015, Padovani2016}. Particle acceleration is achieved through the process of diffusive shock acceleration (DSA), a first-order Fermi mechanism \citep{Drury1983}. Recently, \citet{Purser2016} found that out of 28 high-mass ($M > 8\ \mathrm{M_{\odot}}$) YSOs associated with jets, 14 of them were also associated with nonthermal radio emission, suggesting that this could be a common feature of high-mass YSO jets. In the high-mass YSO jet associated with the Herbig Haro objects HH80 and HH81, \citet{Carrasco-Gonzalez2010} detected linear polarisation  in the jet, showing the presence of synchrotron emission and the orientation of the magnetic field.

DG Tau A is a classical T-Tauri star located in the Taurus Molecular Cloud at a distance of $120.8\pm2.2\ \mathrm{pc}$, based on GAIA data \citep{GaiaCollaboration2016,Bailer-Jones2018}. It is associated with a bipolar outflow and was one of the first YSOs to be associated with an optical jet \citep{Mundt1983}, displaying several shocks and knots along its axis. The optical jet has a position angle of $223\degr$ \citep{Lavalley1997} and an inclination to the line of sight of $37.7\degr \pm 2.2\degr$ \citep{Eisloffel1998}. In the radio, close to the source, the emission is elongated in the direction of the outflow and has a spectral index typical of thermal free-free radiation \citep{Lynch2013}.

Further from the source, the outflow exhibits synchrotron emission. In 2012, it was observed using the Giant Metrewave Radio Telescope (GMRT) at 323 MHz and 608 MHz \citep{Ainsworth2014, Ainsworth2016}. At both frequencies, an extended region of nonthermal emission, hereafter referred to as knot C, was detected $\sim 14''$ southwest of DG Tau A or $\sim 3 \times 10^3\ \mathrm{au}$, slightly offset from the axis of the outflow. This was hypothesised to be associated with the bow shock of the jet. Combined with previous observations from the Karl G. Jansky Very Large Array (VLA) at 5.5 GHz and 8 GHz \citep{Lynch2013}, a spectral index of -0.89 was measured. Assuming equipartition between the energy of the relativistic electrons and protons in the source and the magnetic field energy \citep{Beck2005}, the magnetic field strength was estimated to be 110 $\mu$G. Another region of nonthermal emission, hereafter referred to as knot D, was also detected to the northeast of DG Tau A at 323 MHz \citep{Ainsworth2016}, which was assumed to be associated with the counter-jet.

These two emission knots were later detected again at 6 GHz and 10 GHz using the VLA by \citet{Purser2018}. No proper motion was detected in knot C over 4 years suggesting it may not be associated with a bow shock but instead with a quasi-stationary shock due to dense cloud material drifting into the jet. This could also be the case for knot D.

In this letter, we report on the observation of DG Tau A at 152 MHz using the Low-Frequency Array (LOFAR) \citep{Vanhaarlem2013}. This is the second time that a YSO has been observed using LOFAR. The low-mass YSO T Tau was observed in the same observing campaign and reported on in \citet{Coughlan2017}. In section \ref{sec:observations}, we discuss how the observations were carried out and the data reduction process. In section \ref{sec:results}, we describe the resulting image obtained. In section \ref{sec:discussion}, we then model the low-frequency turnover observed in the spectrum of knot C in the jet of DG Tau A. Finally, we present our concluding remarks in section \ref{sec:conclusions}.

\begin{figure*}
    \centering
    \includegraphics{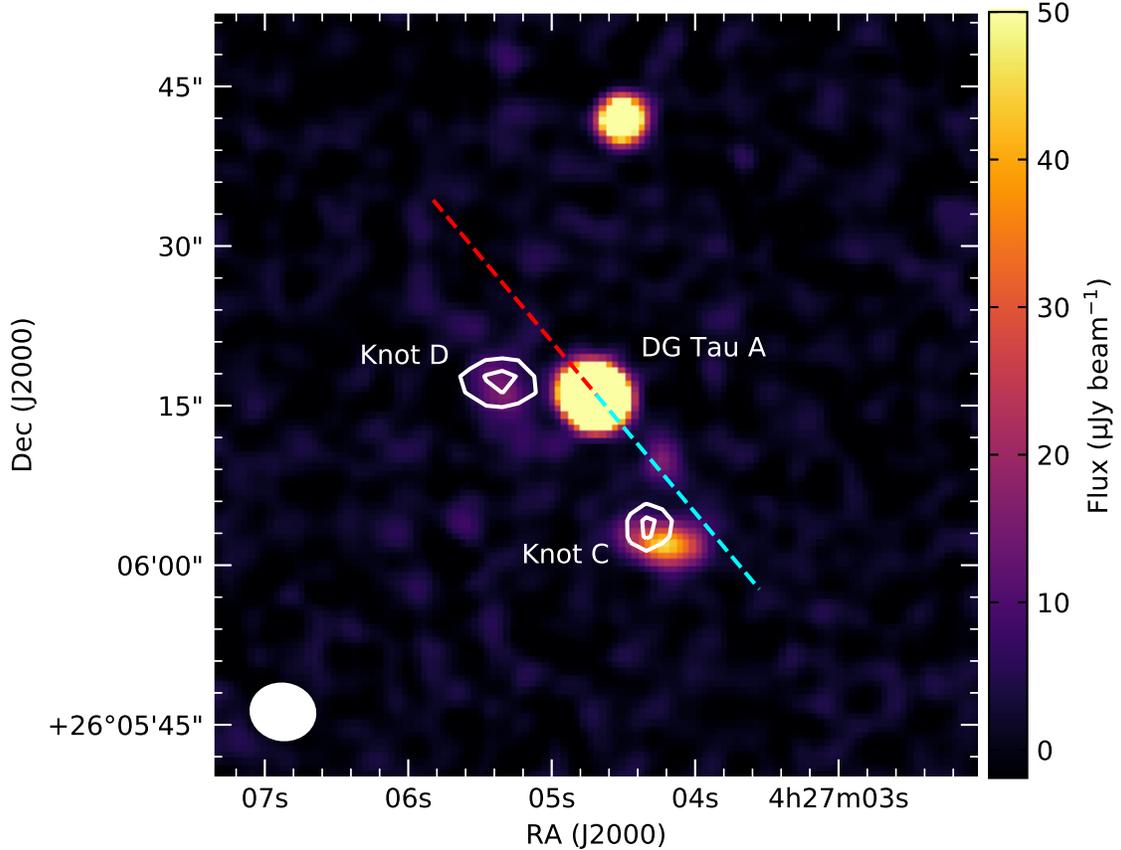}
    \caption{Contour plot of the DG Tau A observation at 152 MHz overlaid on a colour map of a 6 GHz image of DG Tau A from the VLA \citep{Purser2018}. The restoring beam for the LOFAR observation was $6.04^{\prime\prime} \times 5.25''$ with a position angle of $80.7\degr$ and is shown in the bottom left corner. The root-mean-square noise level in the LOFAR observation was $\sigma_{rms} = 90\ \mathrm{\mu Jy\ beam^{-1}}$ and the contour levels are -3, 3 and 4$\ \times\ \sigma_{rms}$. The noise level in the VLA observation was $\sigma_{rms} = 1.9\ \mathrm{\mu Jy\ beam^{-1}}$. The receding and approaching lobes of the jet are indicated by the red and blue dashed lines respectively.}
    \label{DGTau_image}
\end{figure*}

\begin{table*}[t]
    \centering
    \begin{tabular}{cccccc}
        \hline 
        Epoch & Instrument & $\nu$ & $\sigma_{rms}$ & Knot C & Knot D \\
         & & (MHz) & ($\mathrm{\mu Jy\ beam^{-1}}$) & ($\mu$Jy) & ($\mu$Jy) \\
        \hline 
        2013.91 & LOFAR & 152 & 90 & $490 \pm 145$ & $860 \pm 205$ \\ 
        2012.95 & GMRT & 323 & 127 & $1350 \pm 207$ & $936 \pm 196$ \\ 
        2012.95 & GMRT & 608 & 80 & $1087 \pm 175$ & -- \\ 
        2015.10 & VLA & 2500 & 8 & $219 \pm 29$ & $122 \pm 34$ \\
        2015.10 & VLA & 3500 & 5 & $159 \pm 23$ & -- \\
        2012.22 & VLA & 5400 & 6.2 & $73 \pm 8$ & -- \\ 
        2016.15 & VLA & 6000 & 1.9 & $95 \pm 6$ & $42 \pm 4$ \\
        2012.29 & VLA & 8500 & 8 & $39 \pm 9$ & -- \\ 
        2016.10 & VLA & 10000 & 2.7 & $72 \pm 15$ & -- \\
        \hline 
    \end{tabular}
    \caption{Knot C and knot D flux density measurements}
    \label{Emission_Knot_Fluxes}
\end{table*}

\section{Observations and Data Reduction} \label{sec:observations}
We observed DG Tau A on 2013 November 30 - December 1 (epoch 2013.91), using LOFAR (Project code: LC1\_001). Total on source time was 8 hours with the High Band Array (HBA) using 74 MHz of bandwidth, and a central frequency of 152 MHz. 

Before we received the data, the data were run through the pre-processing pipeline by the ASTRON Radio Observatory. This automatically flagged the data for RFI using \textsc{AOFlagger} \citep{Offringa2012} and then averaged in time to 4 s and in frequency to 50 kHz per channel. For LOFAR, there are important direction-dependent effects (DDEs) present due to the ionosphere and imperfect knowledge of the beam shapes, therefore a technique known as facet calibration \citep{VanWeeren2016b} is required to accurately calibrate the data.

Initially, direction-independent calibration was performed using the LOFAR pre-facet calibration pipeline, known as \textsc{Prefactor}. Amplitude and phase gain calibration was performed using the calibrator source 3C147 using the flux density scale from \citet{Scaife2012}. This source was observed in two 10 minute runs, bookending the observations of the target field. The phase solutions were then separated into the contributions due to clock offsets of the individual stations and due to the total electron content (TEC) of the atmosphere. Both the amplitude and clock solutions were transferred from the calibrator to the target field. However, the TEC solutions were not transferred as these are direction-dependent and so they are not applicable to the target field. \textsc{Prefactor} then performed an initial direction-independent phase calibration using a model of the target field obtained from the TIFR GMRT Sky Survey (TGSS) \citep{Intema2017}. Using \textsc{WSClean} \citep{Offringa2014}, each 2 MHz band was then imaged and the clean models produced were then subtracted from the visibilities and converted into sky models. This results in a series of 2 MHz bands with all the sources subtracted and a corresponding sky model for each band.

To account for the DDEs, the LOFAR facet calibration pipeline, known as \textsc{Factor}, was used. This divides the field up into `facets' each of which is centred on a bright, compact source or group of sources which acts as a calibrator. The calibrators are selected based on their brightness and angular size. A brightness threshold of 0.1 Jy and a maximum size of $2^{\prime}$ was used to select the individual sources. Sources which had a separation of less than $9^{\prime}$ were grouped together as a calibrator group. Sources, or groups of sources, whose total combined flux density was brighter than 0.2 Jy were then used as the calibrators. For each facet, the calibrator sources were added back to the data and the data were phase shifted to the position of the calibrator field. The calibrator field was then imaged and several rounds of self-calibration were performed, until the noise was no longer significantly improving. The other sources from the sky model in the entire facet were then added back to the data and the facet was imaged with the solutions for the calibrator field applied to the whole facet. After imaging the facet, an improved sky model was extracted from the image and subtracted from the data with the improved DDE solutions applied. This process was repeated for all of the facets in order of decreasing brightness of the calibrator, with the exception of the facet containing the target which was processed last to ensure that all the artifacts from other facets were removed before the target facet was calibrated.

The bright outlier source 3C123 was also subtracted from the data using the \textit{outlierpeel} task in \textsc{Factor}. A model of 3C123 was obtained using an observation from the LOFAR Long-Term Archive (LTA). The data from this observation were calibrated using \textsc{Prefactor} and then imaged using \textsc{WSClean}.  A model of 3C123 was then extracted from this image using the \textsc{PyBDSF} \citep{Mohan2015} source extraction software and used to calibrate and subtract the source from the DG Tau A observation.

Parts of the frequency spectrum were flagged due to poor amplitude solutions found by \textsc{Prefactor}, likely due to RFI. The final half hour of the observation on all stations, and the entire last hour on station RS407HBA, was also flagged due to poor calibration solutions found by \textsc{Factor}.

After the calibration was complete, all of the sources for each facet were added back to the data and each facet was re-imaged. Final imaging was done using the Briggs weighting scheme \citep{Briggs1995} with a robust parameter of -0.5.

Due to inaccuracies in the LOFAR HBA beam model \citep{VanWeeren2016b, Shimwell2016}, there are large uncertainties in the flux scale. To account for this, the program \textsc{topcat} \citep{Taylor2005} was used to compare the integrated flux densities of several compact bright sources in the field of the LOFAR image with their flux  densities in the TGSS survey. The average ratio of LOFAR flux densities to TGSS flux densities was found to be 1.24. Therefore, the flux densities measured in the LOFAR image were corrected by this factor to account for this. An absolute flux density calibration error for LOFAR of 15\% was then assumed for all the measured flux densities, as used in other LOFAR HBA observations \citep{Shimwell2016, Savini2018}.

Due to the ionosphere there can be large astrometric errors present in LOFAR images. Therefore, the positions of the bright compact sources in the field were compared with their positions in the TGSS survey. A relatively large systematic offset was found of $1.88'' \pm 1.50 ''$ with a position angle for the offset of $11\degr \pm 54\degr$.

\section{Results} \label{sec:results}
After calibrating the visibilities and reducing the data, a radio image of DG Tau A was produced (see Fig. \ref{DGTau_image}).

To measure the flux densities in the image, the task \emph{imfit} in the Common Astronomy Software Application (\textsc{CASA}) \citep{McMullin2007} was used. This measures the integrated flux density by fitting a Gaussian to the emission. The error in the flux density measurements was taken to be a combination of the root-mean-square noise $\sigma_{rms}$ of the image around DG Tau A, the fitting error $\sigma_{fit}$ from the Gaussian fit to the emission and the absolute flux calibration error, which was 15\% for LOFAR: $ \sigma_{S_{\nu}} = \sqrt{\sigma_{rms}^2 + \sigma_{fit}^2 + (0.15 \times S_{\nu})^2}$.

\begin{figure*}
    \centering
    \epsscale{1.17}
    \plottwo{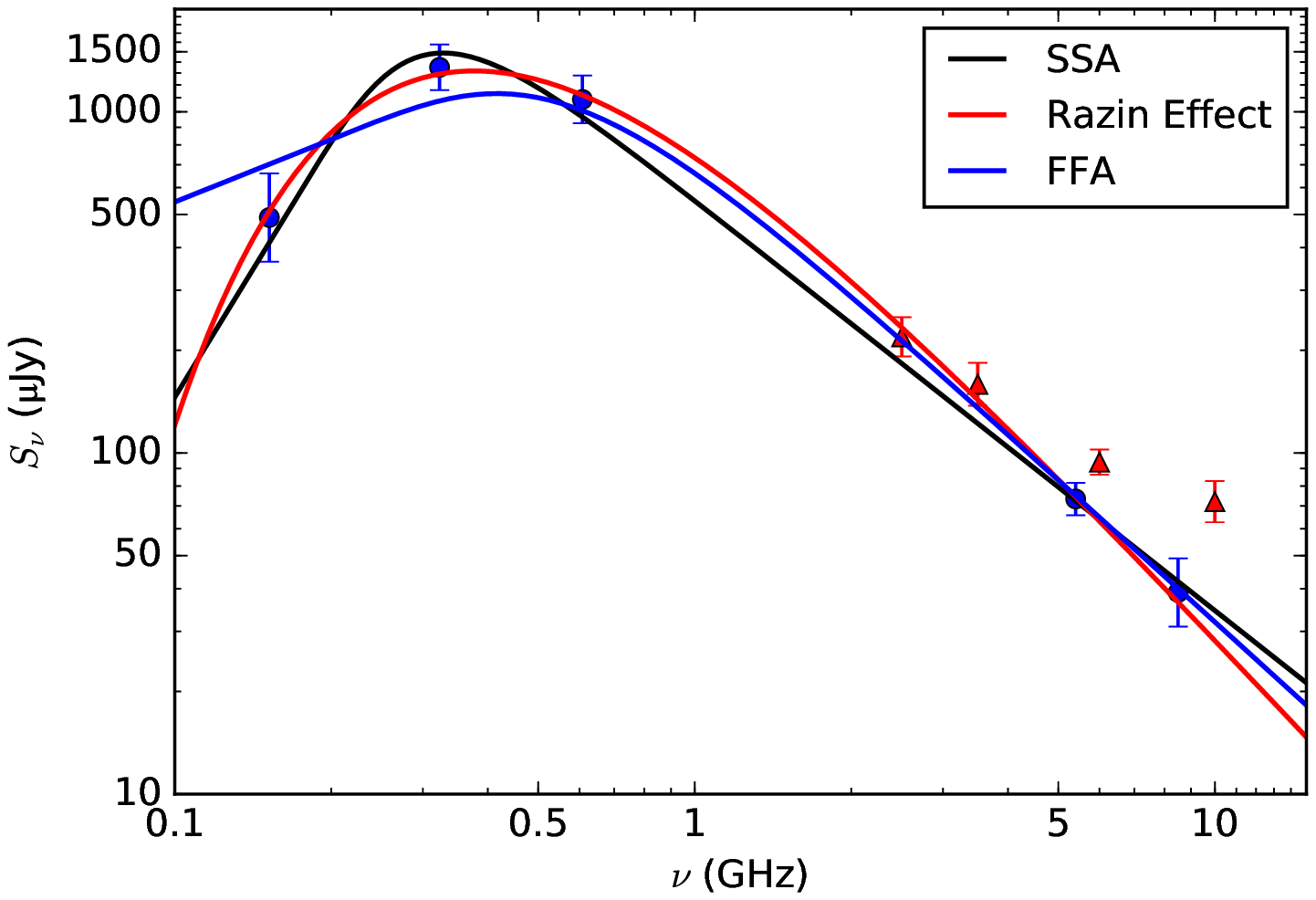}{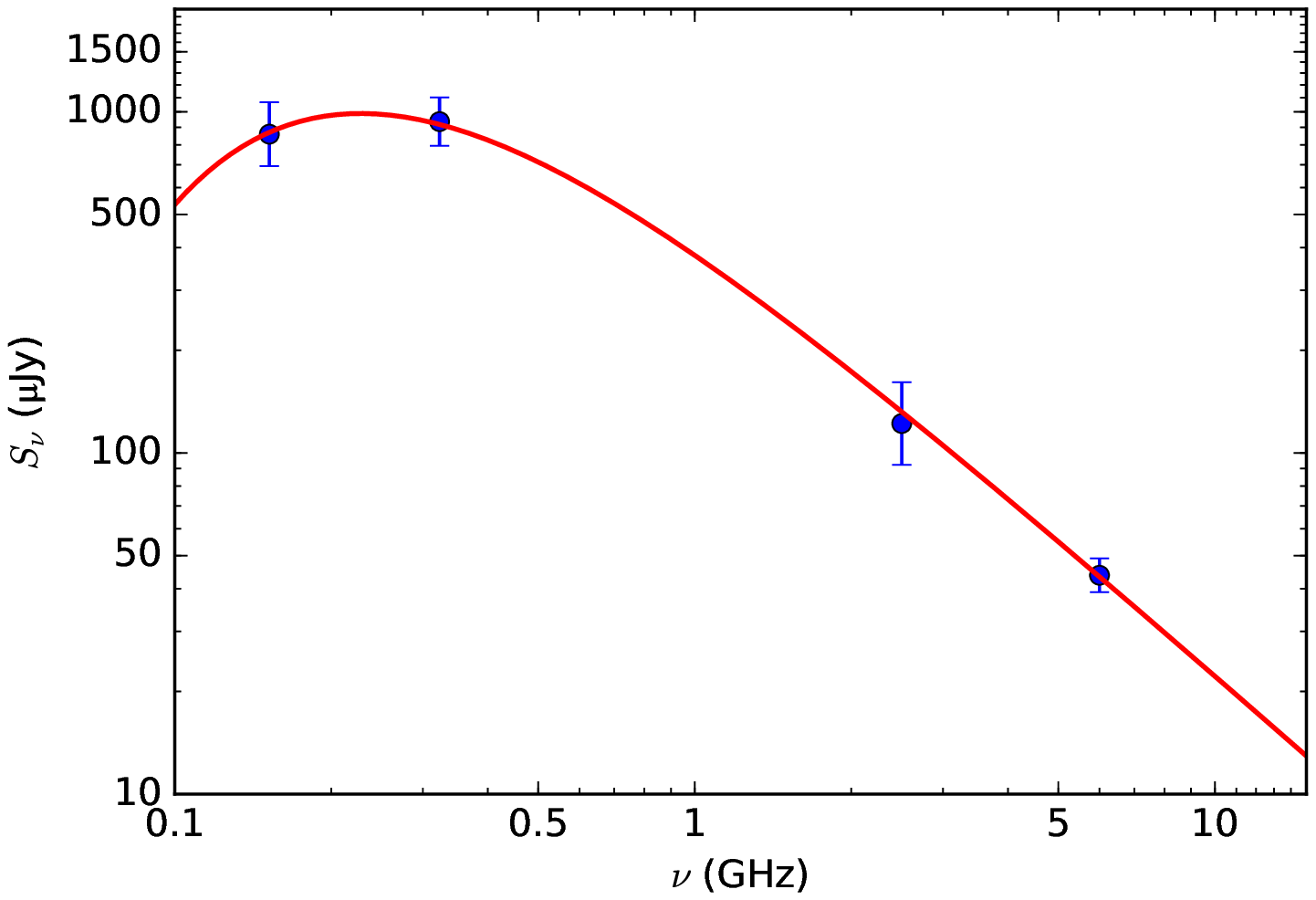}
    \caption{\textbf{Left}: Flux density values of knot C plotted against frequency. Different turnover mechanisms were fitted to the data from LOFAR at 152 MHz, GMRT at 323 MHz and 608 MHz \citep{Ainsworth2014} and the 2012 VLA observations at 5.5 GHz and 8.5 GHz \citep{Lynch2013}, which are indicated by the blue circles. Flux densities from 2015 VLA archival data at 2.5 GHz and 3.5 GHz and from previously published 2016 VLA data at 6 GHz and 10 GHz \citep{Purser2018} are also plotted, indicated by the red triangles, However, these were not fitted as they are from 2015 and 2016 and too different in time to be comparable. \textbf{Right}: Flux density values of knot D plotted against frequency. Data were from LOFAR at 152 MHz, GMRT at 323 MHz \citep{Ainsworth2014}, 2015 VLA archival data at 2.5 GHz and previously published 2016 VLA data at 6 GHz \citep{Purser2018}. A Razin effect turnover has been fitted to the data.}
    \label{spectra}
\end{figure*}

Both knot C and knot D were detected with flux densities of $490\pm145\ \mathrm{\mu Jy}$ and $860\pm205\ \mathrm{\mu Jy}$ respectively. These measurements were combined with previous GMRT and VLA measurements \citep{Ainsworth2016, Lynch2013, Purser2018} and VLA archival data to give the spectra in Fig. \ref{spectra} and are also listed in Table \ref{Emission_Knot_Fluxes}. Unfortunately, the thermal jet was not detected as it is too weak at these frequencies, although we can place a $3\sigma$ upper limit on its peak flux density of $ \lesssim 270\ \mathrm{\mu Jy}$.

The image in Fig. \ref{DGTau_image} appears to show a difference in the position of the peak of the emission for knot C between the LOFAR image and the VLA image at 6 GHz. The peak position in the LOFAR image was offset compared to the VLA image by $2.39''$ with a position angle of $31\degr$. This is within the limits of the systematic offset between the LOFAR image and the TGSS survey suggesting that the difference in position is most likely an astrometric error due to the systematic offset in the LOFAR image.

The integrated flux densities for the GMRT observations \citep{Ainsworth2014, Ainsworth2016} at 323 MHz and 608 MHz were remeasured using \emph{imfit} so that a consistent method to measure the flux densities was used for comparison purposes. At 608 MHz, due to the curved shape of the emission at this frequency, it was necessary to use two Gaussians to accurately model the flux density. To ensure that this approach was accurate, the model was subtracted from the image. The residual image was then inspected to check that essentially all the emission from the source had been removed. The errors were calculated with the same procedure as for the LOFAR measurements, except with a flux density calibration error of 5\%.

In the case of knot C, the emission is clearly much lower than predicted for nonthermal emission (see Fig. \ref{spectra}). Only limited flux density variability was found at higher frequencies, where there was a $\sim 2\sigma$ increase in flux density at 5.5 GHz and 8.5 GHz between 2012 and 2016 \citep{Purser2018}, making it unlikely that there was a significant decrease in flux density between observations. This suggests that a spectral low-frequency turnover has been observed.

\section{Discussion} \label{sec:discussion}
\subsection{Finding the Low-Frequency Turnover Mechanism of Knot C}
To find the low-frequency turnover mechanism responsible, models for each potential mechanism were tested and fitted to the spectrum (see Appendix A). Only the measurements from 2012 and 2013 were considered in order to avoid any issues with flux variability with time.

At first, the most obvious explanation would appear to be synchrotron self-absorption (SSA) \citep{Rybicki1979}. Synchrotron emission is accompanied by absorption, as any radiation emission mechanism must be, where photons interact with electrons in the magnetic field and are absorbed. This depends on the magnetic field strength $B$ and size of the emitting region. Since the emitting region is unresolved, we take the beam size $\theta \approx 5''$ as an upper limit. SSA seems to provide a good fit to the data and gives a turnover frequency of 299 MHz. However, in order to fit the spectrum, it would require a magnetic field strength in the knot of $B =  4.86 \times 10^{14}\ \mathrm{G}$, a highly unrealistic value and in complete disagreement with the equipartition value previously derived ($B = 110\ \mathrm{\mu G}$). While the size of the emitting region used is only an upper limit, it would need to be many orders of magnitude smaller to give realistic values for $B$. This seems unlikely given that at higher frequencies, knot C has been resolved and found to have a deconvolved size of $ 5.3\arcsec \times 1.3\arcsec$ \citep{Purser2018}. Therefore, another explanation is clearly required.

Free-free absorption (FFA) \citep{Rybicki1979} in the emission knot could also provide an explanation. This would depend on the electron temperature $T_e$, electron density $n_e$ and size of the absorbing region. From optical emission line ratio observations \citep{Oh2015}, we assume $T_e \sim 5000\ \mathrm{K}$ and the size of the emitting region was again taken to be the beam size as an upper limit ($\theta \approx 5''$). However, in order to fit the spectrum, an electron density of $n_e \approx 1.2 \times 10^{4}\ \mathrm{cm^{-3}}$ would be required, much higher than the values seen from emission line ratios of $n_e \sim 500\ \mathrm{cm^{-3}}$ \citep{Oh2015}. If the region is smaller than assumed, this would require an even higher electron density. Given this, FFA in the emission knot seems unlikely. While FFA along the line of sight is also possible, this also seems improbable given that knot D exhibits a significantly shallower low-frequency turnover, and so this would require the column density along the line of sight to vary between these two regions.

Another possible explanation is a low-energy cut-off in the energy distribution of the electrons resulting in a low-frequency cut-off in the radiation spectrum. Below the frequency cut-off, the radiation will have a spectral index of $+1/3$ or less, since this is the low-frequency spectral index of radiation from a single electron \citep{Rybicki1979} so the spectral index cannot be steeper than this. However, this would not provide a good fit to the data, since the spectral index below the low-frequency turnover is much steeper than $+1/3$.

Finally, the low-frequency turnover could be due to the Razin effect \citep{Rybicki1979}.  For synchrotron radiation, the emitting source is a plasma, and so the refractive index, $n_r$, is less than 1. Therefore, the phase velocity of light in the plasma becomes $c/n_r$, with $n_r$ decreasing towards low frequencies. This suppresses the beaming effect responsible for synchrotron radiation at low frequencies and therefore the synchrotron emission is reduced. The turnover frequency for the Razin effect, or Razin frequency, $\nu_R$, is given by \citet{Ginzburg1965}:
\begin{equation}
    \nu_R = 20 \frac{n_e}{B}\ \mathrm{MHz}
\end{equation}
where $n_e$ is the electron density in $\mathrm{cm^{-3}}$ and $B$ is the magnetic field strength in $\mu$G. The Razin effect was found to provide a good fit to the data and gives a turnover frequency of $\nu_R = 629\pm30\ \mathrm{MHz}$.  From the \ion{S}{2} emission line ratio \citep{Oh2015}, the electron density in the jet at $14\arcsec$ from the source is known to be $n_e \sim 500\ \mathrm{cm^{-3}}$. Based on this, and the turnover frequency derived, a magnetic field strength of $B \sim 20\ \mathrm{\mu G}$ was calculated.

By calculating the constant $\kappa$ in the electron energy distribution (see Appendix A) and integrating over the relevant range of electron energies, where the minimum and maximum electron energies were the energies of electrons emitting at the minimum and maximum frequency observed, the density of relativistic electrons in the source can be estimated. For the Razin effect fit, for an angular size of $5''$ and using the values of $B$ and the electron power law index $p$ derived from the fit, it was found to be $ n_e^{\mathrm{rel}} \sim 5 \times 10^{-6}\ \mathrm{cm^{-3}}$.

Based on all the possibilities considered, the Razin effect appears to be the most likely explanation for the low-frequency turnover, as it is the only explanation which provides a realistic spectral fit and plausible physical parameters. It should be noted that regardless of whether the Razin effect is responsible, this calculation still provides a lower limit on the magnetic field strength as if it were lower, the spectrum would already have turned over at a higher frequency due to the Razin effect.

The optically thin spectral indices calculated for each of the fits were $-1.20\pm0.06$, $-1.39\pm0.29$ and $-1.66\pm0.04$ for SSA, FFA and the Razin effect respectively. These values for $\alpha$ are significantly steeper than the value measured by \citet{Ainsworth2014}, which was $-0.89\pm0.07$, and also significantly steeper than the maximum value expected for a synchrotron source ($\alpha \approx -1$), unless spectral ageing is involved. However, it should be noted that there are only two points at higher frequencies (5.5 GHz and 8.5 GHz), which are quite close in frequency space, meaning that it is hard to be certain about the accuracy of the estimates for the spectral index of the optically thin region of the spectrum. The data points for the 2015 and 2016 VLA data seem to show a shallower spectral index, but as mentioned previously these were not included in the fit to avoid potential issues with variability.

\subsection{Comparing with Previous B Estimates}
Despite the importance of the magnetic field to the jet launching mechanism and collimation, gauging the strength of the magnetic field in YSO jets has proven very difficult. As a result, measurements of the magnetic field strength are scarce. Therefore, if the Razin  effect is responsible for the turnover, this would provide a valuable measurement of the magnetic field strength along a YSO jet. In optical jets, the pre-shock magnetic field strength has previously been estimated through its effect on the post-shock compression (the ratio of the post-shock electron density to the pre-shock electron density). For the distant bow shocks in the jets of two less evolved YSOs \citep{Morse1992, Morse1993}, which are more embedded within their surrounding envelopes, $B$ was estimated to be $\sim 20 - 30\ \mathrm{\mu G}$ for $n_e \sim 100 - 200\ \mathrm{cm^{-3}}$ at distances of $\sim 5 - 6 \times 10^4\ \mathrm{au}$ from the YSO. This compares well with the measurement for DG Tau A of $\sim20\ \mathrm{\mu G}$ at $\sim 3 \times 10^3\ \mathrm{au}$ from the source. The previously mentioned HH80-81 jet by comparison, had a larger field strength of $\sim 200\ \mathrm{\mu G}$, $ \sim 1 \times 10^5$ au from the source, measured using the equipartition method, although this is for a massive YSO, so the conditions in the jet may be very different.

\subsection{Equipartition Magnetic Field Strength}
The equipartition magnetic field was also recalculated based on the spectrum observed in this paper (see Appendix B). Using the flux value measured at 608 MHz, a typical spectral index for a synchrotron source of $\alpha \approx -0.5$, a volume for the source of $V \approx 4 \times 10^{41}\ \mathrm{m^{3}}$ and a filling factor of $f = 0.5$, an equipartition magnetic field value of $B_{eq} \approx 690\ \mathrm{\mu G}$ was obtained, larger than the previous value calculated in \citet{Ainsworth2014} of $110\ \mathrm{\mu G}$.

A value for the equipartition magnetic field was also obtained based on the Razin effect fit. By equating the energy density in the nonthermal particles in the source with the magnetic field energy density, an expression for $B_{eq}$ dependent on the parameters for the Razin effect fit can be obtained (see Appendix B). Using this method, an equipartition magnetic field value of $B_{eq} \approx 376\ \mathrm{\mu G}$ was obtained, again larger than the value from \citet{Ainsworth2014}, although smaller than the value calculated with standard equipartition magnetic field formula in the previous paragraph. From this value we can also obtain an estimate for the nonthermal electron density in the equipartition regime by again integrating over the relevant range of electron energies. This gives an equipartition nonthermal electron density of $ n_e^{\mathrm{rel}} \sim 2 \times 10^{-7}\ \mathrm{cm^{-3}}$.

Though the value for $B$ from equation (1) from the Razin effect turnover frequency is significantly smaller than the equipartition values calculated, this is not unreasonable given the large uncertainties associated with the equipartition method, particularly regarding the volume of the source. In addition, it is not certain that the source components should be close to equipartition.

\subsection{Knot D}
Unfortunately, knot D was not detected at 608 MHz with the GMRT or in the 2012 VLA observations. However, an approximate spectrum was plotted including VLA archival data from 2015 at 2.5 GHz and with the VLA observation at 6 GHz from 2016. It seems to tentatively show a low-frequency turnover in its spectrum (see Fig. \ref{spectra}). A Razin effect turnover was fitted to the spectrum and appears to be a plausible explanation for the low-frequency turnover in this emission knot as well. However, detections of this knot at only 4 frequencies, which are separated in time, make it difficult to be certain and so no detailed calculations were carried out. The spectrum appears to have a lower turnover frequency than knot C. If this low-frequency turnover is also caused by the Razin effect, this could indicate different magnetic field strengths or electron densities in the two emission knots. More sensitive observations at other frequencies are required for any detailed analysis.

\section{Conclusions} \label{sec:conclusions}
In this letter, we have successfully detected DG Tau at 152 MHz using LOFAR. This is only the second time that a YSO has been detected at such low frequencies and the first time that LOFAR has detected nonthermal emission from a YSO. We found a low-frequency turnover in the synchrotron spectrum of an emission knot of the jet, the first time such a turnover has been detected in the jet of a YSO. Considering the possible mechanisms for the turnover, synchrotron self-absorption (SSA) can almost certainly be dismissed as a possibility given the implausible magnetic field strength required. Alternatively, the Razin effect seems to provide the best explanation as it provides the most realistic physical parameters for the emission knot.

If the Razin effect is the correct mechanism, we can then obtain an estimate for the magnetic field in the emission knot of $ B \sim 20\ \mathrm{\mu G}$ at a projected distance of $\sim 3 \times 10^3$ au from the central source. This would be the first time that the magnetic field strength along a YSO jet has been measured based on the low-frequency turnover in a synchrotron spectrum. Though it should be noted that the magnetic field strength in the jet could differ from that in the emission knot due to effects such as amplification and shock compression. In the future, this method could be applied to other YSO jets to measure the magnetic field strength at locations along the jet, although it requires the presence of nonthermal emission and knowledge of the electron density.

Unfortunately, there are currently no other known low-mass YSOs associated with nonthermal emission, though it has been detected in several high-mass YSOs. However, recent improvements in the sensitivity of radio interferometers such as the VLA,  and the existence of sensitive low-frequency interferometers such as LOFAR may change this. In addition, the next generation of ultra-sensitive interferometers, such as the Square Kilometre Array (SKA) and the Next Generation Very Large Array (ngVLA), could allow very sensitive studies of nonthermal emission to be carried out. Detecting the low-frequency turnover of synchrotron radiation may therefore prove to be a valuable method of measuring the magnetic field and complement other methods such as polarization measurements.\\

A. F.-J., S. J. D. P. and T. P. R. would like to acknowledge support from the European Research Council advanced grant H2020--ERC--2016--ADG--74302 under the European Union's Horizon 2020 Research and Innovation programme.

We would like to thank the anonymous referee for providing constructive comments and helping to improve the quality of this manuscript.

This paper is based (in part) on data obtained with the International LOFAR Telescope (ILT) under project code LC1\_001. LOFAR \citep{Vanhaarlem2013} is the Low Frequency Array designed and constructed by ASTRON. It has observing, data processing, and data storage facilities in several countries, that are owned by various parties (each with their own funding sources), and that are collectively operated by the ILT foundation under a joint scientific policy. The ILT resources have benefited from the following recent major funding sources: CNRS-INSU, Observatoire de Paris and Universit\'e d'Orl\'eans, France; BMBF, MIWF-NRW, MPG, Germany; Science Foundation Ireland (SFI), Department of Business, Enterprise and Innovation (DBEI), Ireland; NWO, The Netherlands; The Science and Technology Facilities Council, UK; Ministry of Science and Higher Education, Poland.
\software{Prefactor \citep{VanWeeren2016b}, 
        Factor \citep{VanWeeren2016b}, 
        AOFlagger \citep{Offringa2012},
        WSClean \citep{Offringa2014},
        PyBDSF \citep{Mohan2015}, 
        Topcat \citep{Taylor2005},
        CASA \citep{McMullin2007}, 
        Matplotlib \citep{Hunter2007}, 
        APLpy \citep{Robitaille2012}}

\appendix

\section{Modelling the low-frequency turnover}
To see which mechanism could best explain the low--frequency turnover, models were fitted to the spectrum. An isotropic, homogenous, spherically symmetric, emitting region was assumed for each of the models with a uniform, randomly oriented magnetic field.

For a synchrotron emitting region, the  electron  energy distribution can be described by a power law of the form \citep{Rybicki1979}:
\begin{equation}
    N(E) dE = \kappa E^{-p} dE 
\end{equation}{}
The flux density produced can be obtained from the equation of radiative transfer:
\begin{equation}
    S(\nu) = \Omega \frac{j_{\nu}}{\alpha_{\nu}} \left( 1 - e ^{-\alpha_{\nu} l}  \right)
\end{equation}
where $\Omega$ is the solid angle of the emitting region, $j_{\nu}$ is the emissivity coefficient, $\alpha_{\nu}$ is the absorption coefficient and $l$ is the size of the emitting region

For synchrotron emission, the emissivity coefficient is given by:
\begin{equation}
    j_{\nu} \propto \kappa B^{(p+1)/2} \nu^{-(p-1)/2}
\end{equation}
where $B$ is the magnetic field strength and $\nu$ is the frequency of the emission. The absorption coefficent for SSA is given by:
\begin{equation}
    \alpha_{\nu}^{\mathrm{SSA}} \propto \kappa B^{(p+2)/2} \nu^{-(p+4)/2}
\end{equation}
while the absorption coefficient for FFA at low frequencies is given by:
\begin{equation}
    \alpha_{\nu}^{\mathrm{FFA}} \propto T^{-3/2} n_e^2 \nu^{-2}
\end{equation}
where $T$ is the electron temperature and $n_e$ is the electron density.

Since the physical parameters in these models are coupled, when fitting the spectrum, combined parameters were used. For SSA, the parameters fitted were $p$, $P_1$ and $P_2$ where $P_1 \equiv \Omega B^{-1/2}$ and $P_2 \equiv \kappa B^{(p+2)/2} l$. From the value for the combined parameter $P_1$, calculated from the spectrum fit, the magnetic field strength required in the source for SSA to occur can be calculated if the angular size of the source is known:
\begin{equation}
    B = \left( \frac{\Omega}{P_1} \right)^2
\end{equation}

Similarly, for FFA, $p$, $P_3$ and $P_4$ were fitted where $P_3 \equiv \kappa \Omega B^{(p+1)/2} T^{3/2} n_e^{-2}$ and $P_4 \equiv T^{-3/2} n_e^2 l$. From the combined parameter $P_4$, the electron density in the source required for FFA to occur can be calculated if $T$ and $l$ are known:
\begin{equation}
    n_e = \sqrt{ \frac{P_4 T^{3/2}}{l} }
\end{equation}

For the Razin effect, which is a reduction in the emission as opposed to an absorption effect like FFA or SSA, the low-frequency turnover has a form where below the cut-off frequency $\nu_R$, the flux density decreases exponentially with frequency and it has a noticeable effect above $\nu_R$, e.g. at $10\nu_R$ there is still a 10\% reduction in flux \citep{Hornby1966}. Therefore, the low-frequency turnover can be approximated by multiplying the optically thin spectrum by a factor of $e^{-\nu_R / \nu}$ \citep{Dougherty2003}. The parameters fitted are a constant of proportionality $K$, $\nu_{\mathrm{R}}$ and $p$.

$K$ is related to the constant $\kappa$ in the electron energy distribution in equation (A1) by the equation:
\begin{equation}
    K = \frac{2.344 \times 10^{-25}}{4\pi} (1.253 \times 10^{37})^{(p-1)/2} a(p)  \Omega l B^{(p+1)/2} \kappa
\end{equation}
where $a(p)$ is a constant dependent on $p$, obtained from \citet{Longair2011}. If $\kappa$ is obtained from the value for $K$ for the fit to the spectrum, an estimate for the density of relativistic electrons can then be obtained by integrating over the range of relevant electron energies. The energy $E$ of an electron whose peak emission intensity is at frequency $\nu$ is given by the equation:
\begin{equation}
    E = \left( \frac{\nu}{CB} \right)^{1/2}
\end{equation}
where $C = 1.22 \times 10^{10} / (m_e c^2)^2$. Therefore, the minimum and maximum electron energies can be taken to be the energies corresponding to the minimum and maximum frequencies observed.

The fitting parameters for all of the turnover mechanisms fitted are listed in Table \ref{Fitting_parameters}, as well as the corresponding values of the optically thin spectral index $\alpha$, which is related to $p$ by the relation $\alpha = - (p - 1) / 2$.

\begin{table}[h]
    \centering
    \begin{tabular}{ c c c c }
        \multicolumn{4}{c}{SSA: } \\
        \hline
        $P_1$ & $P_2$ & $p$ & $\alpha$ \\
        $(\mathrm{T^{-1/2}})$ & $( \mathrm{J^{p-1}\ m^{-2}\ T^{(p+2)/2}})$ & & \\
        \hline
        $(8.06\pm2.01) \times 10^{-15}$ & $(2.03\pm7.45) \times 10^{-23}$ &  $3.41\pm0.12$ & $-1.20\pm0.06$ \\ 
        \hline
        \\
        \multicolumn{4}{c}{FFA: } \\
        \hline
        $P_3$ & $P_4$ & $p$ & $\alpha$ \\
        $(\mathrm{J^{p-1}\ m^3\ T^{(p+1)/2}\ K^{3/2})}$ & $(\mathrm{K^{-3/2}\ m^7)}$ & & \\
        \hline
        $(4.08\pm77.68) \times 10^{-71}$ & $(3.63\pm3.84) \times 10^{28}$ & $3.78\pm0.57$ & $-1.38\pm0.29$ \\
        \hline
        \\
        \multicolumn{4}{c}{Razin effect: } \\
        \hline
        $K$ & $\nu_R$ & $p$ & $\alpha$ \\
        $(\mathrm{W\ m^{-2}\ Hz^{(p-3)/2}})$ & MHz & & \\
        \hline
        $(1.22\pm1.10) \times 10^{18}$ & $625\pm32$ & $4.32\pm0.08$ & $-1.66\pm0.04$ \\ 
        \hline
    \end{tabular} 
    \caption{Fitting parameters for each of the turnover mechanisms fitted. Note that the units for some of the parameters depend on the electron power law index $p$. The optically thin spectral indices $\alpha$ corresponding to the values of $p$ derived are also listed.}
    \label{Fitting_parameters}
\end{table}

\section{Equipartition magnetic field strength}
The standard formula for the equipartition magnetic field strength is based on the magnetic field corresponding to the minimum total energy in the source required to generate the synchrotron radiation observed and is given by \citet{Longair2011} as:
\begin{equation}
    B_{eq} = \left[ \frac{3 \mu_0}{2} \frac{G(\alpha) (1+k) L_{\nu}}{Vf} \right]^{2/7}
\end{equation}
where $\mu_0$ is the vacuum permeability, $G(\alpha)$ is a constant dependent on $\alpha$ and the minimum and maximum observed frequencies of the radio spectrum, $k$ is the ratio of the energy of relativistic protons in the source to that of relativistic electrons in the source (typically $\approx 40$ for electrons undergoing acceleration in a non-relativistic shock \citep{Beck2005}), $L_{\nu}$ is the luminosity of the source at frequency $\nu$, $V$ is the volume of the source and $f$ is a filling factor describing the fraction of the volume occupied by emitting material.

Alternatively, the equipartition magnetic field can be estimated by equating the energy density in the nonthermal particles in the source $U_{\mathrm{NT}}$ and the energy density of the magnetic field $U_{\mathrm{B}}$ so that $U_{\mathrm{NT}} = U_{\mathrm{B}}$, where:
\begin{equation} 
    U_{\mathrm{NT}} \approx (1 + k) \frac{\kappa}{p-2} E_{\mathrm{min}}^{-p+2}
\end{equation}
where $E_{\mathrm{min}}$is the electron energy corresponding to the minimum frequency observed in the spectrum, and:
\begin{equation}
    U_{\mathrm{B}} = \frac{B^2}{2 \mu_0}
\end{equation}
The parameters $\kappa$ and $E_{\mathrm{min}}$ are both dependent on $B$. By substituting the expressions for these parameters and then rearranging $U_{\mathrm{NT}} = U_{B}$, we can obtain an expression for the equipartition magnetic field strength:
\begin{equation}
    B^{7/2}_{eq} = \frac{82 \mu_0}{p-2} \frac{5.36\times10^{25}}{(1.253\times10^{37})^{(p-1)/2}} \frac{K}{a(p) \Omega l} \left( \frac{\nu_{\mathrm{min}}}{C} \right) ^{(-p+2)/2}
\end{equation}
Using the values for $K$ and $p$ from the Razin effect fit, a value for the equipartition magnetic field strength can be obtained.

\section{Testing spectrum fits}
Each of the fits for the low-frequency turnover mechanisms fitted to the spectrum of knot C was tested using the reduced chi-squared test to check the accuracy of the fit. This is given by:
\begin{equation}
    \chi_{\nu}^{2} = \frac{1}{k} \sum_{i=1}^{N} \frac{(y_i - f(x_i))^2}{\sigma _i^2}
\end{equation}
where $y_i$ are the measured data values, $f(x_i)$ are the theoretical data values based on the parameters of the fit, $\sigma_i$ are the errors for the data values and $k$ is the degrees of freedom which is given by $k = n - m$ where $n$ is the number of data values and $m$ is the number of fitted parameters. A value of $\chi_{\nu}^{2} \approx 1$ indicates a good fit while $\chi_{\nu}^{2} \gg 1$ indicates a poor fit. The values calculated for each fit are listed in Table \ref{Chi_squared_table}.

\begin{table}[h]
    \centering
    \begin{tabular}{c c}
        \hline
        Turnover Mechanism & $\chi_{\nu}^2$ \\ 
        \hline
        SSA & 0.65 \\ 
        FFA & 2.05 \\
        Razin effect & 0.11 \\ 
        \hline
    \end{tabular} 
    \caption{$\chi_{\nu}^2$ values for each of the fitted low-frequency turnover mechanisms}
    \label{Chi_squared_table}
\end{table}

\end{document}